\documentstyle{laa}
\begin{document}


\thesaurus{08(02.01.1; 09.03.2; 02.04.2; 08.19.4; 09.19.2; 02.19.1)}
\title{Cosmic Rays}
\subtitle{III. The cosmic ray spectrum between $\bf 1\;\rm\bf GeV$ and
$\bf 10^{\bf 4}\;\rm\bf GeV$\\ and the radio emission from supernova remnants}
\author{Peter~L.~Biermann\inst{1} \and Richard~G.~Strom\inst{2}}
\offprints{Peter~L.~Biermann}
\institute{Max Planck Institut f\"ur Radioastronomie, Auf dem H\"ugel 69,
 D-5300 Bonn 1, Germany
\and
 Netherlands Foundation for Research in Astronomy, Radiosterrenwacht,
 P.O.Box 2, NL-7990 AA Dwingeloo, The Netherlands}
\date{Accepted for publication, March 1993}

\maketitle
\begin{abstract}

Based on a conjecture about the diffusion tensor of relativistic particles
perpendicular to the magnetic field at a shock, and considering particle
drifts, we develop a theory to account for the cosmic ray spectrum between $1
\; \rm GeV$ and $10^4 \; \rm GeV$.  The essential assumption is that the mean
free path perpendicular to the magnetic field is independent of energy and has
the scale of the thickness of the shocked layer.  We then use the basic concept
that the cosmic ray particles are accelerated in a supernova shock that travels
through the interstellar medium.  Radio observations, with additional
supporting evidence from optical and X-ray data, suggest that small additional
and transient bowshocks in locally free expansion contribute to the injection,
while it is the overall shock which accounts for injection as well as the main
acceleration of energetic particles, and we obtain a reasonable picture of the
physical process with such a concept. Physically important ingredients besides
the presence of a strong shock are diffusion, drifts, convection, adiabatic
cooling, the injection history, and the topology of the magnetic field, here
assumed for simplicity to be homogeneous in the interstellar medium.  The
result
is a spectrum, which for strong shocks in a gas with adiabatic index $5/3$
yields a spectrum of $E^{-2.42}$.  Interstellar turbulence with a Kolmogorov
spectrum then leads by leakage from the galactic disk to a spectrum which is
$E^{-2.75}$, as observed in cosmic rays, and as deduced from radio observations
of the nonthermal emission from our Galaxy as well as that of all other
well-observed galaxies. We argue that the ratio of cosmic ray electrons to
protons is determined by the amount of expansion which takes place from the
cessation of electron injection to the break-up of the shell by cooling
instabilities. Since the highest particle energy reached derives from
geometrical arguments, it depends on the charge of the nucleus and so higher
$Z$ elements are predicted to reach higher energies.

\end{abstract}

\section{Introduction}

The origin of cosmic rays is still not completely understood.  There are a few
well-accepted arguments:

\noindent a) The cosmic rays below about $3 \; 10^4 \; \rm GeV$ are due to the
explosions of stars into the normal interstellar medium (Lagage \& Cesarsky
1983);

\noindent b) The cosmic rays from near $10^3$ GeV up to the knee, at $5 \; 10^6
\; \rm GeV$, are very likely due to explosions of massive stars into their
former stellar winds (V\"olk \& Biermann 1988). The consequences of this
concept have been checked by calculating the cosmic ray abundances and
comparing them with observations (Silberberg et al. 1990); the comparison
suggests that up to the highest energy where abundances are known, this concept
successfully explains the data. It is especially interesting that no mixing
from the supernova ejecta is required to account for the known cosmic ray
abundances, Wolf Rayet winds and other strong stellar winds around evolved
stars as sources are all that is needed at present;

\noindent c) For the energies beyond the knee there is no consensus; Jokipii
\& Morfill (1987) argue that a galactic wind termination shock might be able
to provide these particles, while Protheroe \& Szabo (1992) propose an
extragalactic origin, although in either case the matching of the flux at the
knee from two different source populations remains somewhat problematic;

\noindent d) For the cosmic rays beyond the ankle, at about $3 \; 10^9 \; \rm
GeV$ an extragalactic origin is required because of the extremely large
gyroradii of such particles.
\par
In the first paper of this series (Biermann 1993, hereafter paper CR I) we have
proposed that a major effect to consider is the transport coefficient in a
geometry where the magnetic field is perpendicular to the shock normal,
commonly referred to as a perpendicular shock, and where the shock is
spherical. Then we investigate the propagation of a shock wave in either a
stellar wind or into a homogeneous medium. In such a case particle drifts at
the shock and in the upstream and downstream regions are important (see, e.g.,
Jokipii 1987).
\par
Observations can be a guide here: The explosions into the interstellar medium
are explosions where the magnetic field is nearly perpendicular to the shock
direction over most of $4 \pi$ steradians. Radio polarization observations of
supernova remnants clearly indicate what the typical local structure of these
shocked plasmas is.  The observational evidence (Milne 1971, Downs \& Thompson
1972, Reynolds \& Gilmore 1986, Milne 1987, Dickel et al.\ 1988b) has been
summarized by Dickel et al. (1991) in the statement that all shell-type
supernova remnants less than $1000$ years old show dominant radial structure in
their magnetic fields near their boundaries.  There are several possible ways
to explain this: Rayleigh-Taylor instabilities between ejected and swept-up
material can lead to locally radial differential motion and so produce a
locally radial magnetic field (Gull 1973).  It could be due to strong radial
velocity gradients of various clumps of ejecta, or result from clouds being
overrun that now evaporate and cool the surrounding material (all of which are
mentioned by Chevalier [1977]).
\par
It may seem paradoxical that we will be considering perpendicular shocks where
the observations show a parallel magnetic field, but it must be remembered that
the observed field is a superposition of the configurations of all emitting
regions in the telescope beam. The degree of polarization in all of the objects
referred to above is low (typically 10\%
from a perfectly uniform field (generally 65\%
being both parallel and perpendicular components present. In young shells the
former dominate, presumably reflecting the importance of the radial motions
involved. (Although not relevant to the polarization, it might be noted that
the parallel components will cancel on average, while the perpendicular
components do not since they derive from the preexisting magnetic field
structure.)
\par
The radial magnetic field configuration is also found in a few shell remnants
somewhat older than $1000$ yr, such as RCW 86 (SN 185) and Pup A (with a
kinematic age of some 3700 yr, Winkler et al. 1988), showing that radial motion
in some form persists beyond the youthful stage. In most ``mature" remnants,
however, the field is tangential, or complex (Milne 1987). While a
radially-directed magnetic field provides the most compelling evidence for
large-scale differential motions parallel to the shock direction, there are
other indications as well. It is generally believed that the oxygen-rich fast
moving knots (fmk's) in Cas A are ejecta from the progenitor (Kamper \& Van den
Bergh 1976). Oxygen-rich knots have also been seen in Pup A (Winkler et al.
1988) and the similarly aged G 292.0+1.8 (Braun et al. 1986), and may also be
present in a number of extragalactic SNRs (Winkler \& Kirshner 1985).
Radially-directed radio structures have been mapped in Cas A, and their motion
has also been determined (Braun et al. 1987). Recent Rosat observations of the
older ($10\,000$ yr) Vela SNR show similar features in the X-ray emission which
have apparently penetrated the main SNR shell (plenary talk by J. Tr\"umper
presented at the Texas/Pascos '92 Symposium, December 1992) and are believed to
result from fmk-like ejecta. (Their very soft X-ray spectra would make similar
features difficult to detect in more distant remnants.)
\par
The important conclusion for us here is that there appear to be strong radial
differential motions in perpendicular shocks which provide the possibility that
particles get convected parallel to the shock direction.  We assume this to be
a diffusive process, and note that others have also pointed out that diffusion
may be the key to understanding shock acceleration (e.g.\ Falle 1990). Our task
here will be to derive a natural velocity and a natural length scale, which can
be combined to yield a diffusion coefficient. A classical prescription is the
method of Prandtl (1925):  In Prandtl's argument an analogy to kinetic gas
theory is used to derive a diffusion coefficient from a natural scale and a
natural velocity of the system.  Despite many weaknesses of this generalization
Prandtl's theory has held up remarkably well in many areas of science far
beyond the original intent.  We will use a similar prescription here.
\par
Consider the structure of a layer shocked by a supernova explosion
into a homogeneous medium in the case that the adiabatic index of the gas is
$5/3$ and the shock is strong.  Then there is an inherent scale in the system,
namely the thickness of the shocked layer, in the spherical case for a strong
shock limit $r/12$.  There is also a natural velocity scale, namely the
velocity
difference of the flow with respect to the two sides of the shock.  Both are
the smallest dominant scale in velocity and in length; we will use the
assumption that the smallest dominant scale is the relevant scale again to
derive diffusion coefficients and other scalings.
\par
In paper CR I we introduced the basic theory on the basis of a transport
coefficient as described above, and in paper CR II (Biermann \& Cassinelli
1993) we tested it with the observations of nonthermal radio emission from
accelerated particles in shocks traversing stellar winds, while in paper CR IV
(Stanev et al.\ 1993) we checked the spectrum and chemical composition against
air shower data in the particle energy range from $10^4$ GeV to $3 \, 10^9$
GeV. The interpretation of the radio observations of stars before and after
they explode led us to the concept that there is a critical shock velocity of
order $10^9$ cm s$^{-1}$ above which electrons get injected into the
acceleration process in a fashion similar to the nuclei, and below which this
injection is severely restricted, an effect of several orders of magnitude. One
possible choice for this critical velocity is that at which the post-shock
thermal electrons reach relativistic velocities, or, in equipartition, the
whistler velocity becomes relativistic.
\par
Biermann (1992), Rachen (1992), Rachen \& Biermann (1992, 1993: paper UHE CR I)
have proposed that the ultrahigh energy particles arise from hot spots in
nearby radio galaxies; this hypothesis leads to a successful and nearly
parameter-free explanation (Rachen 1992) of the intensity, chemical composition
and spectrum of these particles (see Rachen et al. 1993, paper UHE CR II) with
the important proviso that the mean free path in intergalactic space should be
not much smaller than the characteristic distances between the sources and us,
and may be similar to the scale of the large-scale bubbles in the universe.
These results independently confirm the existence of a steep spectrum component
of heavy nuclei, identified with galactic sources in paper CR IV.
\par
The observations of the cosmic rays themselves, and of the nonthermal radio
emission from our Galaxy as well as from all other well-observed galaxies
strongly suggest that in all carefully studied galactic environments, the
cosmic rays have a universal spectrum of very nearly $E^{-8/3}$ below the knee
at $5 \,10^6$ GeV (below about $10$ GeV for the energetic electrons); direct
air shower experiments show the spectrum beyond the knee to be
well-approximated by $E^{-3}$.  This overall spectrum is clearly influenced by
propagation effects, since particles at different energies have different
probabilities of escaping from the disk of the galaxy. It appears to be a
reasonable hypothesis to approximate the interstellar turbulence spectrum by a
Kolmogorov law, which leads to an interstellar diffusion coefficient
proportional to $E^{1/3}$.  Such an energy dependence then requires a source
spectrum of Cosmic Rays of approximately $E^{-7/3}$ below the knee, and
approximately $E^{-8/3}$ above the knee.
\par
In this paper we propose to derive such a spectrum for particle energies below
about $10^4$ GeV.  The basic hypothesis is, again, that we consider explosions
into a homogeneous interstellar medium.  We will use the standard Sedov
solution to the expansion of such an explosion into the interstellar medium
(e.g. Cox 1972).

\section{Definition of the task}

We consider the acceleration of particles in a shock that propagates into a
homogeneous magnetic interstellar medium.  The acceleration of particles is
governed by the standard theory (Parker 1965), which includes the effects of
first, time change, then radial diffusion, latitude diffusion, radial drift,
latitude drift, then compression and finally sources (see paper CR I). The
orientation which we use is such that the symmetry axis is parallel to the
prevailing magnetic field, and we assume that to first order the magnetic field
structure is unidirectional and homogeneous in the undisturbed interstellar
medium. We take the cosine $\mu$ of the colatitude as our coordinate along
$\theta$. The interstellar magnetic field is known to contain both systematic
ordered components and a random component (Beck 1986) of similar strength;
hence
we will demonstrate below (Sect.\ 8) that the simplification of using a uniform
field does not introduce any unnecessary restrictions.  The components of the
diffusion tensor of interest here are the radial diffusion term $\kappa_{rr}$
and the latitude diffusion term $\kappa_{\theta\theta}$.
\par
The Sedov similarity solution for an explosion into a homogeneous medium can be
written as a function of shock radius $r$ and shock speed $U_1$; as a function
of time $t_4$, in units of $10^4$ years; explosion energy $E_{51}$, in units of
$10^{51}$ ergs; and density of the interstellar medium $n_{\rm o}$ in units of
cm$^{-3}$. This gives (Cox 1972)

\begin{equation}
r\;=\;13.7 \,t_4^{2/5}\,(E_{51}/n_{\rm o})^{1/5} \,{\rm pc} \end{equation}
and

\begin{eqnarray}
U_1/c\; & = & \; 1.8\,10^{-3}\,t_4^{-3/5}\,(E_{51}/n_{\rm o})^{1/5}\;
\nonumber\\
& = & \; 0.090\,({r/{\rm pc}})^{-3/2}\,(E_{51}/n_{\rm o})^{1/2}.
\end{eqnarray}
Among the modifications which can be made to Eqs.\ (1) and (2) with practical
application to SNRs, it is possible to account for (radial) density gradients
in both the ambient medium and the ejecta. Chevalier (\cite{cheva}) suggests,
for example, that a modification which may be appropriate for Tycho would
result in $r\propto t^{4/7}$ instead of the $2/5$ power-law in Eq.\ (1).
Kinematic investigations of Tycho (Strom et al.\ \cite{str:gos}), Kepler
(Dickel et al.\ \cite{dic:sau}) and SN 1006 (Long et al.\ \cite{lon:bla}) show
that while there are deviations from the ideal $2/5$ power-law, they are not
substantial, and it is moreover unknown what their precise cause is. We shall
consequently adopt the Sedov solution for the sake of simplicity and because it
seems to be a fair representation of reality.
\par
Our boundary conditions are the usual:  We inject particles at some low energy
which is assumed to be independent of all relevant properties of the problem,
i.e. not dependent on radial distance, magnetic field strength or latitude.
Downstream we assume the flow to take particles out of the system with the
normal probability $4\,U_2/c$ (see, e.g., Drury 1983), where $U_2$ is the
downstream velocity relative to the shock.
\par
We propose to derive the essential properties of the particle distribution
function by analytical means, using heuristic arguments. The key will be the
form of the diffusion tensor, especially the radial component $\kappa_{rr}$.

\section{Diffusion perpendicular to the average magnetic field}

Our basic conjecture (see CR I), then, is that the convective random walk of
energetic particles perpendicular to the magnetic field can be described by a
diffusive process due to fast convective motions with a downstream diffusion
coefficient $\kappa_{rr,2}$ which is given by the thickness of the shocked
layer together with the velocity difference across the shock, and is thus
independent of energy:

\begin{equation}\kappa_{rr,2} \;=\;{1 \over 9}\, {U_2 \over U_1}\, r \, (U_1
\, - \,U_2) \end{equation}
Here we have assumed that in the combination of transport coefficients
perpendicular and parallel to the magnetic field (see Sect.\ 8) the
perpendicular convective transport dominates. As is obvious from Eqs.\ (51) and
(52) below, this is a very good approximation for nearly all angles over the
$4\pi$, since the parallel diffusion coefficient is linear with $E$, and the
perpendicular convective transport is independent of particle energy. The exact
limiting condition is $(1-\mu^2)/\mu^2 \gg E/E_{max}$, which defines the very
small solid angle over which the parallel diffusion is stronger.
\par
The upstream diffusion coefficient is then obtained (again, see paper CR I) by
using the same velocity scale and the downstream length scale multiplied by the
density ratio, which gives $r/3$ for a strong shock.  Again, the gyroradius of
the same maximum particle energy as in the shocked layer gives the same scale.
Since the magnetic field is lower by a factor of $U_2/U_1$ upstream, the
gyroradius of the maximum particle energy upstream is also $r/3$.  Hence the
upstream diffusion coefficient is

\begin{equation}\kappa_{rr,1} \;=\; {1 \over 9}\, r \, (U_1 \, - \, U_2)
\end{equation}
It immediately follows that

\begin{equation}{\kappa_{rr,1} \over {r \, U_1}}\;=\;{\kappa_{rr,2} \over
{r \, U_2}}\;=\; {1\over 9} \, \left(1 \, - \, {U_2 \over U_1}\right)
\end{equation}
\par
For these diffusion coefficients, it also follows that the residence times
(Drury 1983) on both sides of the shock are equal and are

\begin{equation}{4 \, \kappa_{rr,1} \over U_1 c}\;=\;{4 \,\kappa_{rr,2} \over
U_2 c}\;=\;{4\over 9}\, {r \over c} \, \left(1 \, - \, {U_2 \over U_1}\right).
\end{equation}
Adiabatic losses then cannot limit the energy reached by any particle since
they run directly with the acceleration time, both being independent of energy,
and so the limiting size of the shocked layer limits the energy that can be
reached to that where the gyroradius just equals the thickness of the shocked
layer.  This then leads to a potential maximum energy of

\begin{equation}E_{\rm max}\;=\; {1 \over 3}\,{U_2 \over U_1}\, Z e r B_2
\;=\; {1 \over 3}\,Z e r B_1 \end{equation}
where $Z e$ is the particle charge and $B_{1,2}$ is the magnetic field strength
on the two sides of the shock. This means, once again, that the energy reached
corresponds to the maximum gyroradius the system will allow. It also means that
we push the diffusive picture right up to its limit where the diffusive scale
becomes equal to the mean free path and the gyroradius of the most energetic
particles. We will derive a more restrictive condition below (Sect.\ 8).

\section{Particle Drifts}

Consider particles which are either upstream of the shock, or
downstream; as long as the gyrocenter is upstream we will consider the
particle to be there, and similarly downstream.
\par
In general, the energy gain of the particles will be governed primarily by
their adiabatic motion in the electric and magnetic fields.  The expression
for the energy gain is well known and given in, e.g., Northrop (1963, equation
1.79), for an isotropic angular distribution where one term comes from the
drifts and a second from the induced electric field (see paper CR I).  We
explicitly work in the shock frame,  separate the two terms and consider the
drift term first. The second term is accounted for further below, in Sect.\ 5.
\par
The $\theta$-drift can be understood as arising from the asymmetric component
of the diffusion tensor, the $\theta r$-component.  The natural scales there
are the gyroradius and the speed of light, and so we note that for (Forman et
al. 1974)

\begin{equation}\kappa_{\theta r}\;=\;{1 \over 3} \,r_{\rm g} \,c ,
\end{equation}
the exact limiting form derived from ensemble averaging, we obtain the drift
velocity by taking the proper covariant divergence (Jokipii et al. 1977); this
is not simply (spherical coordinates) the $r$-derivative of $\kappa_{\theta
r}$. The general drift velocity is given by (see, e.g., Jokipii 1987)

\begin{equation}V_{{\rm d},\theta}\;=\;c \,{E \over {3 Z e}}\,{\rm
curl}_{\theta} {{\bf B} \over B^2} . \end{equation}
The $\theta$-drift velocity is thus zero upstream and only finite downstream
due to curvature:

\begin{equation}V_{{\rm d},\theta}\;=\;{c \over 3}\,{r_{\rm g}\over r}
\end{equation}
where $r_{\rm g}$ is now taken to be positive (see above).
\par
It must be remembered that there is a lot of convective turbulence which
increases the curvature: The characteristic scale of the turbulence is $r/12$
for strong shocks, and thus the curvature is $12/r$ maximum.  Taking half the
maximum as average we obtain then for the curvature a factor of $6/r$ which is
six times the curvature without any turbulence; this increases the curvature
term by a factor of six thus changing its contribution from $1/3$ to $2$ in the
numerical factor in Eq.\ (10). Hence the total drift velocity is

\begin{equation}V_{{\rm d},\theta}\;=\;{1 \over 2} \,\left({U_1 \over
{U_2}}\right)\, {{c \, r_{\rm g}}\over r} ,
\end{equation}
now written for arbitrary shock strength.  It is easily verified that the
factor in front of Eq.\ (11) is two for strong shocks where $U_1/U_2=4$. With
$\Delta E_1\;=\;0 $ we have then downstream

\begin{equation}
{{\Delta E_2}\over E}\;=\;{2 \over 9}{U_1 \over c}\, \left(1-{U_2 \over
U_1}\right)
\end{equation}
which is the total energy gain from drifts.

Let us consider then one full cycle of a particle remaining near the shock and
cycling back and forth from upstream to downstream and back. The energy gain
just due to the Lorentz transformations in one cycle can then be written as

\begin{equation}\left({\Delta E \over E}\right)_{\rm LT}\;=\;{4 \over 3} {U_1
\over c} \left(1 -{U_2 \over U_1}\right).
\end{equation}
Adding the energy gain due to drifts, Eq.\ (12), we obtain

\begin{equation}{\Delta E \over E}\;=\;{4 \over3} {U_1 \over c} \left(1 -{U_2
\over U_1}\right) x
\end{equation}
where

\begin{equation}x\;=\;1+{1 \over 6}. \end{equation}

\section{Source expansion and particle injection history}
\subsection{Expansion and injection history, shell}

Consider how long it takes a particle to reach a certain energy:

\begin{equation}{dt \over dE}\;=\;\left[8 \;{\kappa_{rr,1} \over U_1 c}\right]
\left[{4 \over 3} {U_1 \over c} \left(1-{U_2 \over U_1}\right) x E\right]^{-1}.
\end{equation}
Here we have used $\kappa_{rr,1} / U_1 \;=\; \kappa_{rr,2} / U_2$. Since we
have, from the Sedov solution [Eqs.\ (1) and (2)], $r\;=\;(5/2) \,U_1 t $,
this leads to,

\begin{equation}{dt \over t}\;=\;{dE \over E} {{3 U_1} \over {U_1 - U_2}} {2
\over x} {\kappa_{rr,1} \over {r U_1}}\,{5 \over 2} \end{equation}
and so to a dependence of

\begin{equation}t(E)\;=\;t_{\rm o} \; \left({E \over E_{\rm o}}\right)^{\beta}
\end{equation}
with

\begin{equation}\beta\;=\;{{{3 U_1} \over {U_1 - U_2}} {2 \over x}
{\kappa_{rr,1} \over {r U_1}}}\,{5 \over 2}. \end{equation}
Particles that were injected some time $t$ ago were injected at a different
rate, say, proportional to $r^b$. Also, in a $d$-dimensional space, particles
have $r^d$ more space available to them than when they were injected, and so we
have a correction factor which is
\begin{equation}\left({E \over E_{\rm o}}\right)^{-{2 \over 5}(b+d) \beta}.
\end{equation}
The combined effect [Eqs.\ (19) and (20)] is a spectral change by

\begin{equation}{{3 U_1} \over {U_1 - U_2}} {2 \over x} (b+d) {\kappa_{rr,1}
\over {r U_1}}. \end{equation}
We note that the factor $2/5$ from the Sedov expansion drops out again.  For
a detailed argument on the sign conventions used here, see paper CR I.  This
expression can be compared with a limiting expansion derived by Drury (1983;
eq. 3.58), who also allowed for a velocity field.  Drury's expression agrees
with the more generally derived expression given here.  This also describes
then the adiabatic losses, due to the general expansion of the shock layer.
Hence the total spectral difference, as compared with the plane parallel case,
is given by

\begin{equation}{{3 U_1} \over {U_1 - U_2}}\;\left[{U_2 \over U_1} \left({1
\over x} -1\right) \;+\;
{2 \over x} (b+d) {\kappa_{rr,1} \over {r U_1}}\right]. \end{equation}
If this expression is positive the spectral index is steeper.
\par
Furthermore, we have to discuss the effect of the expansion of the material in
the shell, in this context already introduced by Drury (1983).  This expansion
is only that beyond the linear expansion of the overall flow already taken into
account.  We define the velocity for the flow field $U(\xi r)$ as a function of
radius $\xi r$ interior to the shock radius $r$ by

\begin{equation}U(\xi r) \;=\;V(\xi) \, \xi \, U_1.  \end{equation}
We thus have, e.g., $V(1) \,=\, V'(1) \,=\, 3/4$.  This velocity field can be
approximated by

\begin{equation}V(\xi)\;=\;3 \,{{\xi^8 \,+\,1} \over {3 \,\xi^8 \,+ \,5}}.
\end{equation}
As shown by Drury (1983), this velocity field  introduces a steepening
by

\begin{equation}{{3 \,U_1} \over {U_1 - U_2}}\,{{\kappa_{rr,2}} \over {r
U_2}}\,
{V'(1) \over V(1)}, \end{equation}
neglecting drifts and in the limit of very small diffusion coefficient.
However, in our context where we consider a finite shell, we have to use a
non-local approximation. To do this we simply equate the expansion of the hot
interior to the shell with the postshock speed of sound, which is given by

\begin{equation} c_2 \;=\; \left({{2 \gamma} \over {\gamma -1}}\right)^{1/2}
\, U_2 , \end{equation}
where the numerical factor is $ \sqrt{5} \,=\, 2.23608\dots$ for an adiabatic
constant of $\gamma \,=\, 5/3$.  The postshock gas expands at the shock in the
observer's frame at $U_1 - U_2$, which is $3 U_2$ for a strong shock.  The
linear expansion itself already provides $(U_1 - U_2)/12$, which is
$U_2/4$.  Hence the extra expansion over linear is $2.75 \, U_2 \, - \, 2.23608
\, U_2$. Using then finite differences instead of derivatives, we have

\begin{equation}{V'(1) \over V(1)} \; \rightarrow \; {{2.75 U_2 - \sqrt{5} U_2}
\over {3 U_2}} \, {23 \over 24} \, {12} \;=\; 1.970, \end{equation}
instead of unity as in Drury's case, which is $1.97007\dots$ times the term
from the local flow field $V(\xi)$.  Here the factor $23/24$ comes from
averaging the postshock radius from $1$ to $1 -1/12$, and thus describing the
average of $\xi / \Delta \xi$.  Hence we have an extra term from this postshock
adiabatic loss due to the shell expansion of

\begin{equation}{{3 \,U_1} \over {U_1 - U_2}}\,{1 \over x}\,{{\kappa_{rr,2}}
\over {r U_2}}\,{V'(1) \over V(1)} \, 1.970 , \end{equation}
which is then slightly less than double the effect which Drury discusses. For a
strong shock then this particular effect adds $0.563$ to the spectral index.
\par
For strong shocks in stellar winds, for which the natural length scale is
$r/4$, the linear part of the expansion is three times higher and so the
remaining difference to the postshock speed of sound is only $2.25 - \sqrt{5}$,
which gives an effect of only $2 \,10^{-3}$ in the spectral index, justifying a
posteriori our neglect of this effect in papers CR I and CR II.
\par
The exponent $b$ [Eq.\ (20)] describes the injection as a power of the radius;
in a Sedov solution the injection, assumed to be proportional to $\rho U_1^2$,
is given by

\begin{equation}\rho U_1^2\;\sim\;r^{-3}. \end{equation}
Hence we have $b = -3$, so that $b+d = 0$. Thus we have for a pure Sedov
solution, using the velocity field term introduced above, a spectral difference
to the plane parallel case of $0.444$ and thus an injection spectrum of

\begin{equation}E^{-2.420}. \end{equation}
However, we note that this spectrum is valid for electrons obviously only if
they are injected into the acceleration process at all, i.e. if the shock speed
is above the critical velocity for electron injection.  Adding the term for
diffusive losses from the Galaxy the final observable spectrum expected is

\begin{equation}E^{-2.753} ,\end{equation}
very close to what is observed, both in Galactic cosmic rays (protons and
nuclei), as well as through the nonthermal radio emission of other galaxies
(Golla 1989, electrons).
\par
Clearly, there is uncertainty in our method for treating the non-linear flow
field; the various ways of correcting the term Eq.\ (25) all lead to numerical
correction factors very close to $2.0$; above we have taken $1.97$.  Further
non-linear corrections, such as also averaging the non-linear flow field
itself,
lead to spectral indices within $0.04$ of the value given in Eq.\ (30).

\subsection{Expansion and injection history, fast moving knots}

Here we have to note that the radio images of Cas A (Anderson et al.\ 1991)
suggest that there are many small additional bowshocks which in fact may do
some of the particle acceleration (see especially Braun et al. 1987). They
are associated with compact clumps related to the fmk/ejecta (which, in fact,
they may well be), and have travelled outwards essentially undecelerated until
their encounter with the high density material associated with the shock.
Since the main shock has undergone significant deceleration (Tuffs 1986, Braun
et al. 1987), the clumps, expelled with a range of radial speeds, are now able
to overtake it, brightening and hence becoming prominent as they do. Each clump
is the apex of a bowshock which is radially oriented, and as it passes the
shock front it becomes a fresh start into the interstellar medium and so can be
presumed to be in free expansion for as long as it survives, until it
ultimately merges back into the overall general flow.  The observations suggest
that these bowshocks, initiated by small blobs of denser material, move with
velocity constant in time, but different for different blobs; when such blobs
happen to move faster than the present average velocity of the shell, they must
have penetrated it in the past, but when their velocity is just the current
value of $r/t = (5/2) U_1$, then they have just caught up with, and hence
penetrate, the shell now.
\par
The drift velocity is then given by a consideration of free expansion into the
homogeneous medium, and we approximate this free expansion here as a spherical
shell of which only a segment is realized.  Writing for the blob radius $r_{\rm
b}$ and its velocity $U_{\rm b}$ we then obtain for the drift, again using the
argument about convective scales [e.g.\ Eq.\ (11)],

\begin{equation}V_{{\rm d},\theta}\;=\;{1 \over 2}\,{U_{\rm b1} \over U_{\rm
b2}}\, {r_{\rm g} \over r_{\rm b}}\,c.
\end{equation}
The downstream residence time is then

\begin{equation}\tau_{\rm res,2}\;=\;{1 \over 3}\,{r_{\rm b} \over 3}\,{U_{\rm
b2} \over U_{\rm b1}}\,(U_{\rm b1} - U_{\rm b2})\,{4 \over {U_{\rm b2} c}}.
\end{equation}
The energy gain due to drifts is then given by

\begin{equation}
{{\Delta E}\over E}\;=\;{2 \over 9}\,{U_{\rm b1} \over c}\,\left(1-{U_{\rm b2}
\over U_{\rm b1}}\right).
\end{equation}
Now the argument above suggests that $U_{\rm b1}\;=\;(5/2) \;U_1$. Using then
the plausible assumption that $U_{\rm b1}/U_{\rm b2}=U_1/U_2$ we obtain

\begin{equation}
{{\Delta E}\over E}\;=\;{5 \over 9}\,{U_1 \over c}\,\left(1 - {U_2 \over
U_1}\right).
\end{equation}
\par
Adding this now to the Lorentz transformation energy gain (but also with the
increased shock speed) we obtain for the parameter $x$ again the value $x=7/6$
[Eq.\ (15)]. In order to calculate the resulting spectrum we have to find now
the effective diffusion coefficient and express it in terms of the shell
parameters, such as shell thickness of the overall explosion, for strong shocks
$r/12$. For free expansion the parameters $b$ and $d$ are $b+d=3$, and the
expansion is linear so that the velocity field is $V'(\xi)=0$.  Hence we have
to fix the residence time now in order to determine the spectrum. Now the total
scale of the blob-induced convective turbulence is, upstream plus downstream,
$(5/4)\,{r_{\rm b}/3}$. This entire scale will grow until at some point it
reaches the overall shell scale $r/12$; then by virtue of the switch-over in
scales the overall shell scale will begin to dominate and the blob expansion
will cease to be important. This means that
\begin{equation} r_{\rm b}={r\over5} , \end{equation}
and also that

\begin{equation}r_{\rm b} U_{\rm b1} \;=\;{1 \over 2}\,r U_1.\end{equation}
Correspondingly we have to insert $\kappa_{rr,1}/(r U_1)=1/24$ and so obtain
for the spectral difference from the plane parallel case $0.715$, which gives
an injection spectrum of

\begin{equation}E^{-2.715} ,\end{equation}
nearly $1/3$ steeper than the spectrum for the overall shock acceleration, Eq.\
(30), suggesting that these smaller shocks provide just some injection for the
larger shocks. However, the smaller shocks may not only run into a geometric
limit, they are likely to run into an energetic limit even earlier (i.e.\ no
longer have sufficient energy to be in free expansion locally), and so cease to
be important before reaching the limit derived above.
\par
As can be seen from the expression for the spectral index, it depends on the
blob radius, and thus on expansion time.  The spectrum steepens with blob
radius $r_{\rm b}$, which is proportional to expansion time for as long as the
blob is in local free expansion.  The spectrum implied is $E^{-2.114}$ at $0.3
\,r_{\rm b}$, then $E^{-2.715}$ at exactly $r_{\rm b}$, to steepen to
$E^{-4.429}$ at $3 \,r_{\rm b}$. Since energetic particles experience both the
effect of the small blobs as well as that of the overall shock, the overall
spectral index for the blobs is expected to be slightly steeper than the
average (non-blob) spectral index of the shell, consistent with observations.
We thus surmise on energetic grounds that these blobs break up below the
critical blob size given above, and note that their radio emission does in fact
fade rapidly as they pass the shock front (Braun et al.\ 1987), which is
consistent with the breakup requirement. This means that the blobs are likely
to change over to a local Sedov expansion before reaching the critical blob
size;  this then would lead to a cessation of electron injection as soon as the
the shock velocity drops below the critical velocity.
\par
The spectrum derived is the one injected by small spherical blobs in local free
expansion.  This spectrum has more power proportional to the square of the
shock velocity, since the shock velocity  $U_{\rm b1}$ is $5/2$ higher than the
overall shell shock velocity $U_1$, and also the effective area is, of course,
approximately half a sphere, giving another factor of two over the area of a
planar circle on the large shell.  Hence the multitude of these little blobs
puts about one order of magnitude more power per area of the large shell into
the injection than the spherical shell shock directly.  We conclude that the
small blobs driven by the fast moving knots may contribute substantially to the
injection of particles, just as surmised by Braun et al. (1987).
(Parenthetically, it should be noted that the possible role of fmk's in
particle acceleration was first investigated by Scott \& Chevalier [1975] who,
however, assumed a second-order Fermi process in the turbulent fmk wakes.)
\par
We add that these small bow shocks are not the only objects that should have
time-variable radio emission; radio observations suggest that there are also
many nearly stationary magnetic substructures, with likely a magnetic field up
to a factor $5$ above that given by the shock-enhanced interstellar field of
about $20 \,\rm \mu \rm G$.  Such substructures can arise when the overall
expansion overruns preexisting features in the interstellar medium, for which a
straightforward analogy with the solar photosphere would suggest that magnetic
substructures of much stronger magnetic fields might be quite common.
\par
There is one immediate check: Braun et al.\ (1987) and Anderson et al.\ (1991)
emphasize that the large bow shock features with the steepest radio spectral
index have the strongest time variation of their flux density. In our simple
concept this can be interpreted as the consequence of two effects: First, with
the size of the blob or bow shock the region emitting scales as the third power
of the radius of the blob and so steadily increases. Second, with the steeper
spectrum the weight of the particle spectrum is shifted to low particle
energies and so the radio emission is increased. For going from a radio
spectral index of, e.g., $-0.6$ to $-0.7$ the combined effect is about a factor
of $4$, equally caused by both effects.
\par
The spectrum derived is very close to the observed galactic cosmic ray spectrum
at low energies and we propose that this is the mechanism to generate it.

\section{The time evolution of the radio spectral index}

Since the shock velocity is steadily decreasing with time for a Sedov
expansion, we have to consider the effect of particle acceleration through
Alfv\'en wave scattering, commonly referred to as second order Fermi
acceleration. The importance of this mode of acceleration has been appreciated
for some time (e.g., Scott \& Chevalier 1975, Chevalier et al.\ 1976).

The energy gain through this effect can be included in our derivation of
${\Delta E}/E$:

\begin{equation}\left({{\Delta E} \over E}\right)_{\rm2F} \;=\;\left({{\Delta
E} \over E}\right)_{\rm LT}\,\left({v_{\rm A} \over U_1}\right)^2 {\rm const}
\;\sim\;\left({{\Delta E} \over E}\right)_{\rm LT}
\,r^3.\end{equation}
This translates into a time-dependent value of $x$:

\begin{equation}x\;=\;1.167 + (r/r_{\rm s})^3.\end{equation}
This in turn translates into a spectral index for the particle distribution
which gets flatter with time and expansion radius:

\begin{equation}1 + {1.657 \over {1.167 + (r/r_{\rm s})^3}}.\end{equation}
The radio spectral index of the nonthermal synchrotron emission then has a
radius-dependence accordingly of

\begin{equation}{0.842 \over {1.167 + (r/r_{\rm s})^3}}.\end{equation}
This leads to a gradual flattening of the spectral index with $r$. The
quantitative behaviour of this simple approximation is in agreement with the
data shown for the few supernova remnants for which we have adequate distances
or know their individual time evolutions (Anderson et al.\ 1991, Fig. 1).
\par
We note that these effects depend on the ambient density $n_{\rm o}$
such that for a low density environment the Alfv\'en velocity is high and
hence second order Fermi acceleration may then become important. However, we
have to emphasize that second order Fermi acceleration depends on the average
strength of the Alfv\'enic wave field across the entire shell region, and since
that is unknown, we can derive the temporal change of the spectral index
only to within an unknown factor, which might be rather small. Thus we see
two ways to explain the trend for overall spectral index to flatten with
diameter (Berkhuijsen 1986, Anderson et al. 1991):
\par
First, there is the possibility that for larger and thus older remnants the
spectral index may in fact be due to a combination of various components
that may have formed over the long time scale available: 1) HII regions with
their rather flat radio spectrum from free-free emission, 2) a pulsar driven
supernova remnant which also has a flatter radio spectrum, and 3) an underlying
synchrotron spectrum of a shell-type supernova remnant as discussed in this
paper. Furthermore, the supershells often found in other galaxies require the
energy input of many stellar winds and/or supernovae and might be an analogy
to the extremely large supernova remnants detected in our Galaxy, and not just
attributable to a single event.
\par
Second, one could argue that only a few remnants survive to the age where
second order Fermi acceleration becomes important, and so these remnants are
not significant for the typical injection spectrum for the release of cosmic
rays into the interstellar medium.  The distribution of supernova remnants is
given by a $r^{+1/2}$ power law in equal logarithmic bins of observed radius,
for any given environmental density $n_{\rm o}$.  The adiabatic expansion
covers a range of only $8.1 \, n_{\rm o}^{-0.12}$ from free expansion to
cooling and break up, and thus this range is rather insensitive to the
density.  However, the breakup radius depends as $n_{\rm o}^{-5/11}$ on
density and so the typical range of at least a factor of $100$ in the
interstellar medium density (see, e.g., Fig. 12 in Berkhuijsen 1986) likely
to be experienced by exploding stars will shift the distribution around by an
extra factor of ten, totally mixing the diagram (see, e.g., Fig. 11 in
Berkhuijsen 1986). Thus, it is close to impossible to disentangle these various
effects all of which lead to selection effects in any observational survey.  We
suspect, however, that it is this second alternative which is more reasonable
in our Galaxy.

\subsection{The time evolution of the radio emission}

Calculating now the radio emission contribution from the shell of
$r\,U_2/(3\,U_1)$, we note that the ram pressure is proportional to $U_1^2$,
which in the Sedov solution [Eq.\ (2)] is proportional to $r^{-3}$.  The total
radio emission from the shell is then constant with time if the efficiency of
injecting electrons, $\eta$, is a constant. The luminosity is then given by

\begin{equation}L_{\nu}\;=\;5.4 \,10^{23}\,B_{-5.3}^{1.710}\, \eta_{-1} \,
E_{51} \,{\nu_{9.0}}^{-0.710}\;{\rm erg\,s^{-1}Hz^{-1}}
\end{equation}
and so does not depend on the interstellar medium density; since the adiabatic
expansion is a condition of constant energy, all such dependences drop out. We
have used here as reference $0.1$ for $\eta$, $5 \, \rm \mu G$ for the
unperturbed interstellar magnetic field strength, and $1$ GHz for the emission
frequency observed. We note that the emission here is coming from a shell of
thickness $r/12$; additional emission further inside might arise if there is a
reverse shock resulting from the transition of free expansion to adiabatic
expansion. Evidence for reverse-shock heating of ejecta may exist in X-ray
images of Tycho (Seward et al.\ \cite{sew:gor}), and has also been
construed from X-ray spectroscopy of SN 1006 (Hamilton et al.\ \cite{ham:sar})
and Tycho (Hamilton et al.\ \cite{hami:sar}).

Furthermore, there is observational evidence that the synchrotron emission
systematically samples the higher magnetic field substructures, by a factor of
up to $5$ and so we can expect the synchrotron emission to be higher by factors
up to $15.7$, leading to an implied radio luminosity of up to $8.5 \,10^{24} \,
{\rm erg\,s^{-1}Hz^{-1}}$, everything else being equal. This luminosity is in
agreement with the upper limit of the distribution of observed luminosities,
suggesting that in their initial phases the efficiency of putting kinetic
energy into an energetic electron population is indeed of order $0.1$.

However, when we derive the normalization factor for the particle spectrum to
use in the approximation that the total energy in all relativistic particles is
$\eta$ times the flow energy, we have implicitly used a constant particle
spectrum.  After having shown how we might understand the early time evolution
of the spectral index, we now have to calculate how this influences the
normalization;  since with an increasingly flatter spectrum the energy in the
particle population shifts to higher values, there are increasingly fewer
particles at lower energy, the very particles responsible for radio emission.
However, this effect is only strong for relativistic particles and is
counterbalanced by the non-relativistic sector of momentum space, so that the
entire effect is only a factor of order $2$ for a change from a synchrotron
spectral index of $-0.7$ to $-0.8$. Therefore, the radio emission is expected
to decrease slowly with time. As soon as electron injection ceases, then simple
aging will decrease the electron spectrum rapidly and Shklovsky's argument
holds (Shklovsky 1968, \S 7), demonstrating how the electron population
decreases simply due to overall adiabatic losses.

\section{The injection into the interstellar medium}

Since the particle spectrum flattens with time and the observed cosmic
ray spectrum corresponds to the early spectral index of Sedov expansion
explosions, there is an apparent discrepancy.  This discrepancy can be resolved
by noting that a Sedov type expansion breaks up when the shell becomes unstable
to cooling;  the observations and our interpretation then suggest that most
supernova remnants break up and release their energetic particle population
into the interstellar medium long before second order Fermi acceleration
becomes relevant.  The Vela supernova remnant may be an example of a
shell that has been breaking up for some time. Spectroscopy has shown that
the motions in Vela are very chaotic (Jenkins et al.\ 1984), with large
departures from simple radial motion, which may be evidence for this.
\par
The evolution of a supernova remnant (see, e.g., for the basic physical
concepts employed Cox 1972) is then divided into four phases:

First, there is free expansion until the interstellar medium overrun by the
explosion shock is of about the same mass as the mass ejected.  This happens at
a radius of

\begin{equation}
R_{\rm e}\;=\;1.92 \,\left({M_{\rm ej} \over M_{\odot}}\right)^{1/3}
\,n_{\rm o}^{-1/3} \;{\rm pc},
\end{equation}
where $M_{\rm ej}$ is the ejected stellar mass and $n_{\rm o}$ is again the
ambient density in cm$^{-3}$.  Numerical simulations demonstrate that the
expansion is noticeably slowed down long before this critical radius is
reached.
\par
Second, we have a Sedov phase with adiabatic expansion and steady fresh
electron injection.  The observations of radio emitting stars suggest that
there is a critical velocity for a shock, below which electron injection is
severely restricted.  The critical speed is of order

\begin{equation}U_{1,\rm crit}\;=\;c\,(m_{\rm e}/m_{\rm p})^{1/2}\;\simeq\;
7000 \;\rm km\,s^{-1} ,
\end{equation}
as suggested by the radio luminosities of OB and WR stars on the one hand, and
radio supernovae on the other (CR II); this comparison suggests just the
approximate velocity, not the actual number, but the particular velocity given
here is consistent with the stellar data. It is highly unlikely that the same
critical velocity would be relevant in the interstellar medium, and so we only
use the concept that there is such a critical velocity and adopt as reference
$1000$ km s$^{-1}$ ($=U_{\rm crit,-2.5}$, in units of $c$) on the grounds that
this would correspond to roughly the same Alfv\'enic Mach number as the
supernova shocks in stellar winds as argued in papers CR I and CR II without
any physical justification at this stage.  We plan to test such a concept with
further stellar examples of nonthermal radio emission.  In our picture the
injection is done by both the bow shocks around the fast moving knots and the
overall shock.  Hence the critical radius is given by

\begin{equation}
R_{\rm crit}\;=\;9.0 \, (E_{51} / n_{\rm o})^{1/3} \, U_{\rm crit,
-2.5}^{-2/3}\; {\rm pc}. \end{equation}
\par
Third, when the expansion velocity has decreased below the critical speed,
the electron injection ceases and we have further adiabatic expansion, but
with the electron population only modified by adiabatic losses.   Adiabatic
losses of a relativistic particle decrease its energy by the ratio of the
initial and final radius scales.
\par
And finally, fourth, we have the break-up phase, which occurs because the
cooling layer is thermally unstable (McCray et al. 1975, Chevalier and Imamura
1982, Smith 1989) leading finally to a break-up of the supernova shell. The
cooling radius is (Cox 1972)

\begin{equation}R_{\rm cool}\;=\;15.6 \;E_{51}^{3/11} \,(10^{22}
L)^{-2/11}\,n_{\rm o}^{-5/11}\;{\rm pc}, \end{equation}
where $L$ is the cooling coefficient; the cooling curve has been compared with
recent detailed calculations by Schmutzler \& Tscharnuter (1993) and is still
a very good approximation at the temperature range of interest here.  During
this time the energy of any individual relativistic electron decreases by
adiabatic expansion as the ratio of the radii from the time when injection
ceases to the time of break up.  Thus the energy density of the electron
population (from conservation of the adiabatic moment only and disregarding
spatial dilution, see below) decreases by the factor ($p=2.420$)

\begin{eqnarray}\left({R_{\rm cool}\over R_{\rm crit}}\right)^{p-1}\!=\;2.18\,
E_{51}^{-0.09}\,n_{\rm o}^{-0.17}\,(10^{22}L)^{-0.26}\,
U_{\rm crit, -2.5}^{0.95} , \end{eqnarray}
showing a very weak dependence on the environmental density $n_{\rm o}$. The
dependence on the critical velocity is very nearly linear. The simple spatial
dilution of the energy density is the same for both protons and electrons and
drops out in the ratio, as long as we are in the Sedov phase (see Shklovsky
[1968], eq. 7.27).  For a tenuous interstellar medium of H-atom number density
about $0.01\rm\;cm^{-3}$, the factor is increased to nearly $4.8$.
\par
This intermediate switch from electron injection with steady acceleration to a
simple adiabatic loss regime will determine the net scaling of the power of the
electron population to that of the proton population in the cosmic rays. The
observations suggest that from $1$ GeV the electron number is only about $0.01$
(Wiebel 1992) of the protons.  From this observed ratio the energy density
ratio
integrated over the entire relativistic part of the particle spectrum, protons
relative to electrons, for the injection spectral index of $-2.42$, is given by
$4.3$. Assuming then that the ratio of electrons to protons is not influenced
by
propagation effects, this suggests an expansion of a factor in radius of order
$3$ between the time when electron injection ceases and the time when cooling
breaks up the shell of the supernova remnant; here we assume that electrons
and protons originally have comparable energy densities of their relativistic
particle populations.
\par
Interestingly, this is very close to the ratio implied above for normal
parameters.  Below, where we argue about the maximum particle energy that can
be attained, we see that the tenuous part of the interstellar medium is likely
to be the phase responsible for the injection.  This is then fully consistent
with our adopted particular value for the critical velocity of $1000$ km
s$^{-1}$ and suggests that we may have identified the physical reason for the
observed electron/proton ratio.

\section{The maximum energy of particles}

The maximum energy particles can possibly reach is given in Sect.\ 3, and
depends  linearly on the magnetic field.  Near the symmetry axis we have
diffusion parallel to the magnetic field, and, given the strong turbulence
induced by the shock we take there the Bohm limit for the upstream diffusion
coefficient:

\begin{equation}\kappa_{\mid \mid}\;=\;{1 \over 3}\,{E \over {Z e B_1}}\,c.
\end{equation}
At the equator we use our radial upstream diffusion coefficient derived
earlier,
here for a strong shock, of

\begin{equation}\kappa_{\perp}\;=\;{{r U_1} \over 12}. \end{equation}
Combining Eqs.\ (49) and (50) for the two upstream diffusion coefficients
(Jokipii 1987) to obtain an acceleration time scale and setting that equal to
the upstream flow time scale of $r/(3 U_1)$ then yields a condition on the
maximum particle energy that can be reached. This condition is

\begin{equation}{ r \over {3 U_1}}\;=\;{4 \over 3}\,{E \over {Z e B_1}}\,{c
\over U_1^2}\, \mu^2 \, + \, {r \over {3 U_1}}\,(1 - \mu^2). \end{equation}
This leads to

\begin{equation}E_{\rm max}\;=\;{1 \over 4}\,Z e B_1 r\,{U_1 \over c}.
\end{equation}
A more detailed consideration of the strength of the magnetic field as a
function of latitude does not change this result.  In addition, since here the
actual coordinate $\mu$ does not appear in the expression, it is obvious that
this condition holds locally over any area of the spherical shock with a
locally homogeneous magnetic field.  Therefore, the overall inhomogeneity of
the magnetic field does not change this result.
\par
Using normal interstellar magnetic fields of $5\;\mu$G, and a standard Sedov
explosion with an arbitrary radius then yields a maximum particle energy of

\begin{equation}E_{\rm max}({\rm protons})\;=\;1.0 \,10^5 \,(r/{\rm pc})^{-1/2}
\,(E_{51}/n_{\rm o})^{1/2}\; \rm GeV  \end{equation}
and

\begin{equation}E_{\rm max}({\rm iron})\;=\;3.0 \,10^6 \,(r/{\rm pc})^{-1/2}\,
(E_{51}/n_{\rm o})^{1/2}\; \rm GeV. \end{equation}
At breakup, when we suggest the actual mixing of the particle population
into the interstellar medium occurs, these maximum energies are

\begin{equation}E_{\rm max}({\rm protons})\;=\;2.6 \,10^4\,F_{\rm CR}\;\rm GeV
,
\end{equation}
and

\begin{equation}E_{\rm max}({\rm iron})\;=\;7.4 \,10^5\,F_{\rm CR} \;\rm GeV.
\end{equation}
where we use

\begin{equation}F_{\rm CR}\;=\;E_{51}^{0.364} \,(10^{22}L)^{0.091}\,
n_{\rm o}^{-0.273}. \end{equation}

The protons ought to dominate due to the normal abundances in the interstellar
medium, and higher energies for protons are clearly reached in the tenuous hot
part of the interstellar gas where the densities are of order $0.01 \,\rm
cm^{-3}$, and so maximum proton energies are possible to about $1.0 \,10^5\,
\rm GeV$. These numbers are similar to earlier estimates of the maximum
particle energy in a Sedov expansion phase of a supernova. In paper CR IV we
obtain from a fit to the air shower data an estimate for this number of $1.2
\,10^5\, \rm GeV$, fully consistent with these arguments, and suggesting that
the cosmic ray injection is most effective in the tenuous part of the
interstellar medium, an argument which has been made before on quite different
grounds.

\section{Summary}

In this third paper of a series on the process of cosmic ray acceleration we
have again used a basic conjecture (Sect.\ 3) on the diffusion of particles in
a shock perpendicular to the magnetic field.  In a similar vein we have
introduced a number of heuristic arguments which require testing against
observations.  We have already tested and further explored the consequences of
this concept in another paper (CR II) on the nonthermal radio emission of
Wolf-Rayet stars and demonstrate that our concept can produce the proper radio
spectral indices, luminosities and temporal behaviour.  In a further
communication we have tested in detail the predictions of this model as regards
the chemical abundances of cosmic rays (CR IV). What we have shown in papers CR
I, II, IV and here, is that the entire cosmic ray spectrum from $1$ GeV out to
$3\,10^9$ GeV, including the knee energy and the chemical abundances, can be
understood with the same physical ingredients already well-tested in the solar
wind shock, namely a strong shock, diffusion, drifts, convection, adiabatic
cooling, and injection history.
\par
In this paper in particular, we have used the concept introduced in paper CR II
of a critical velocity for electron injection to fashion an argument for the
observed electron/proton ratio in cosmic rays.  We have derived an estimate for
the upper particle energy that can be reached in such explosions, of order
$10^5$ GeV, close to what had been argued previously and what is required by
the air shower data (paper CR IV). We have also derived the proper cosmic ray
spectrum, consistent with direct measurements as well as with radio
observations.  It obviously follows that the abundances in such cosmic rays
ought to be close to those of the interstellar medium itself, as is observed.

\acknowledgements {The essential part of this work was carried out during a
five-month sabbatical in 1991 of PLB at Steward Observatory at the University
of
Arizona, Tucson.  PLB wishes to thank Steward Observatory, its director, Dr.
P.A. Strittmatter, and all the local colleagues for their generous hospitality
during this time and during many other visits.  PLB also wishes to thank Drs.
E. Berkhuijsen, D.C. Ellison, T.K. Gaisser, J.R. Jokipii, K. Mannheim, H.
Meyer,
R. Protheroe, S.P. Reynolds, M.M. Shapiro, T. Stanev and R. Tuffs for extensive
discussions of cosmic ray, supernova and stellar physics. High energy physics
with PLB is supported by grants DFG Bi 191/6,7, 9 (Deutsche
Forschungsgemeinschaft), BMFT grant (DARA FKZ 50 OR 9202) and NATO travel grant
(to T. Stanev and PLB) CRG 910072. The research of RGS is supported by the
Netherlands Organisation for Scientific Research (NWO).}

\end{document}